\documentclass[aps,prl,twocolumn,superscriptaddress,reprint]{revtex4-2}
\pdfoutput=1
\usepackage{amsmath}
\usepackage{graphicx}
\usepackage{bbm}
\usepackage{amssymb}
\usepackage{enumitem}
\usepackage{booktabs}
\usepackage{verbatim} 
\usepackage{soul}
\usepackage{mathrsfs}
\usepackage{bm}
\usepackage[linktocpage=true,
colorlinks=true,
urlcolor=blue,
anchorcolor=blue,
citecolor=blue,
filecolor=blue,
linkcolor=blue,
menucolor=blue,
]{hyperref}

\newcommand{\ex}{\mathrm{e}}
\newcommand{\diff}{\mathrm{d}}
\newcommand{\dd}{\mathrm{d}}
\newcommand{\R}{\mathbb{R}}

\newcommand{\C}{\mathbb{C}}

\newcommand{\hook}{\mathbin{\rule[.2ex]{.4em}{.03em}\rule[.2ex]{.03em}{.9ex}}}

\newcommand{\gi}{\mathrm{GI}}

\usepackage{color}

\def\nn{\nonumber}
\newcommand{\ii}{\mathrm{i}}

\newcommand{\Z}{\mathbb{Z}}

\newcommand{\cB}{\mathcal{B}}
\newcommand{\cC}{\mathcal{C}}

\newcommand{\barK}{K}
\newcommand{\barM}{W}

\newcommand{\baranomaly}{\Phi^{(\text{anom})}}

\newcommand{\newsigma}{e}

\newcommand{\bxi}{\bm{\xi}}

\newcommand\Phianom{\Phi^{(\text{anom})}}
\newcommand{\Sigmanew}{\mathscr{F}}

\begin{document}

\title{Airy functions from quantum M-theory}

\author{Pietro Benetti Genolini}
\affiliation{D\'epartement de Physique Th\'eorique, Universit\'e de Gen\`eve, 24 quai Ernest-Ansermet, 1211 Gen\`eve, Suisse}

\author{Florian Gaar}
\affiliation{Mathematical Institute, University of Oxford, Woodstock Road, Oxford, OX2 6GG, U.K.}

\author{Jerome P. Gauntlett}
\affiliation{Abdus Salam Centre for Theoretical Physics, Imperial College, Prince Consort Road, London, SW7 2AZ, U.K.}

\author{Jaeha Park}
\affiliation{Abdus Salam Centre for Theoretical Physics, Imperial College, Prince Consort Road, London, SW7 2AZ, U.K.}

\author{James Sparks}
\affiliation{Mathematical Institute, University of Oxford, Woodstock Road, Oxford, OX2 6GG, U.K.}

\begin{abstract}
\noindent We show that Airy function partition functions for M2-brane theories may be derived from relative equivariant localization of quantum M-theory. The eleven-dimensional Chern--Simons coupling gives the cubic term in the grand potential, while the $X_8$ correction gives the  charge shift. Fixing the M2-brane charge turns the localized M-theory path integral into an Airy integral. In this way we derive the ABJM result, its toric Calabi--Yau generalizations, and gravitational blocks for black holes and other spacetimes, up to the prefactor and non-perturbative corrections. 
\end{abstract}

\maketitle

\enlargethispage{0.25\baselineskip}

\section{Introduction}\label{sec:intro}

Exact computations of observables in supersymmetric quantum field theories have transformed our understanding of strongly coupled dynamics. 
A key development has been supersymmetric localization, which reduces certain protected path integrals to finite-dimensional integrals or matrix models (see \cite{Pestun:2016zxk} for a review). 
These techniques lead to exact results for partition functions, indices and other observables, which in turn provide sharp tests of holography.

The paradigmatic example is the three-sphere partition function of ABJM theory  \cite{Aharony:2008ug}, describing $N$ M2-branes probing $\C^4/\Z_k$. The localized 
 matrix model admits a Fermi gas formulation \cite{Marino:2011eh}, effectively resumming 
the  perturbative large $N$ expansion  into an Airy function~\cite{Fuji:2011km}. 
Substantial evidence has accumulated that the same Airy structure persists
for  ABJM theory deformed by various parameters, 
and also for more general M2-brane theories  associated with toric Calabi--Yau four-fold singularities. 
We refer the reader to \cite{Bobev:2025ltz} 
for recent results, as well as references to earlier literature.

This structure is most naturally expressed in a grand canonical ensemble. Introducing a chemical potential $\mu$ conjugate to the M2-brane charge $N$, the perturbative grand potential has the universal large $\mu$ form
\begin{align}\label{grandpotential}
J(\mu,b,\bxi) & =  \frac{C(b,\bxi)}{3}\mu^3 + B(b,\bxi) \mu + A(b,\bxi)
\nonumber\\
& \quad   
+ \mathcal{O}(\ex^{-\mu})\, .
\end{align}
Here $b$ is a squashing parameter of the three-sphere $S^3_b$ on which
 the Euclidean M2-brane theory is defined, while $\bxi$ 
represents various field theory parameters,
including Chern--Simons levels $k$ and real masses/$R$-charges. 
The partition function is the inverse Laplace transform 
\begin{align}\label{Laplace}
Z(N,b,\bxi)  = \frac{1}{2\pi\ii}\int_{-\ii \infty}^{\ii \infty} \diff \mu \exp\left[ J(\mu,b,\bxi)- N \mu\right]\, ,
\end{align}
which is 
an Airy function up to a constant prefactor and exponentially small non-perturbative corrections:
\begin{align}\label{partitionfunction}
Z = C^{-\frac{1}{3}}\ex^{A} \text{Ai}\left[\frac{(N-B)}{C^{\frac{1}{3}}}\right]\left(1+ \mathcal{O}(\ex^{-\# \sqrt{N}})\right)\,.
\end{align}
This has been derived rigorously 
in some instances, while more generally substantial evidence has been given using  
numerics, saddle point analyses, and other methods \cite{Bobev:2025ltz}.

From the viewpoint of quantum gravity and holography, the result \eqref{partitionfunction}  is striking.
The leading $N^{3/2}$ term in the large $N$ expansion is well-known to be 
captured by classical supergravity, while the universal charge shift $N\mapsto N-B$
that generates an infinite number of $1/N$ corrections 
should encode genuine quantum-gravitational information. 
In this Letter (and \cite{mediumpaper}) we derive this structure  directly from the eleven-dimensional quantum effective action: relative equivariant localization \cite{BenettiGenolini:2026cdw}
 reduces the classical $D=11$ supergravity coupling $C\wedge G\wedge G$ to the cubic term in the grand potential \eqref{grandpotential}, while the one-loop coupling $C\wedge X_8$ gives precisely the linear term $B\mu$. The M-theory action is naturally computed in the ``M2-brane" ensemble, where $\mu$ is fixed \cite{Gautason:2025plx}. Passing to fixed M2-brane charge $N$ adds the Legendre term $N\mu $ to the action, and the remaining M-theory path integral over $\mu$ is precisely the Airy integral~\eqref{Laplace}. Our approach extends straightforwardly to any spacetime geometry, where for black holes we derive an M-theoretic OSV formula \cite{Ooguri:2004zv}. 

\section{M-theory action}

The action of classical $D=11$ supergravity in Euclidean signature is
\begin{align}\label{D11action}
I = \frac{1}{(2\pi)^8\ell_p^9}\int-R *1 + \frac{1}{2}G\wedge *G  
+ \frac{\ii}{6}C\wedge G \wedge G\, ,
\end{align}
where $\ell_p$ is the Planck length, $G=\diff C$ is the M-theory four-form, and $*$ is the Hodge dual. 
Focusing on supersymmetric configurations and taking an analytic continuation of \cite{Gauntlett:2002fz},  
we introduce bilinear forms in the corresponding spinor parameter $\epsilon$:
\begin{align}\label{bilinears}
K^\flat \equiv \overline{\epsilon}\mskip1mu \Gamma_{(1)}\epsilon\, , \quad 
\Omega \equiv \overline{\epsilon}\mskip1mu \Gamma_{(2)}\epsilon\, , \quad 
\Sigma \equiv \overline{\epsilon}\mskip1mu \Gamma_{(5)}\epsilon\, ,
\end{align}
where $\Gamma_{(r)}\equiv \frac{1}{r!}\Gamma_{\mu_1\cdots\mu_r}
\diff x^{\mu_1}\wedge \cdots \wedge \diff x^{\mu_r}$ and $\overline{\epsilon} \equiv \epsilon^T \cC$, with $\cC$ being the charge conjugation matrix. 
In Euclidean signature $\epsilon$ is a complex Dirac spinor, 
and the forms \eqref{bilinears} are in general complex. 
The Killing spinor equation implies  the vector $K$ dual to $K^\flat$ is Killing, 
with $\mathcal{L}_K G = \mathcal{L}_K\Omega=\mathcal{L}_K\Sigma=0$. 
The supersymmetric Killing vector $K$, and  
cohomology 
for the 
equivariant exterior derivative  $\diff_K\equiv \diff - K \hook\mskip2mu$, will play 
an important role \cite{BenettiGenolini:2023kxp}.

As first noticed in \cite{Nekrasov:2021ked}, there is an equivariantly closed completion of the four-form $G$ to the polyform
\begin{align}\label{phigee}
    \Phi^{G}\equiv G+\Omega\,,
\end{align}
with $\dd_K \Phi^{G}=0$.  
Defining the  gauge-invariant polyform
\begin{align}
    \Phi^{(\gi)}\equiv \frac{1}{3}\Phi^G\wedge(* G+\ii \Sigma)\,, 
\end{align}
one verifies using supersymmetry and the equation of motion for $C$ that 
\begin{align}\label{pair}
\diff_K \Phi^{(\text{GI})} = - \Phianom\, , \  \Phianom \equiv \frac{\ii}{6}\Phi^G\wedge \Phi^G\wedge \Phi^G\, .
\end{align}
Imposing also the trace of the Einstein equation
leads to what we call the  $D=11$ Euclidean (partially on-shell) ``supersymmetric action'',
\begin{align}\label{ISUSY}
I_{\text{SUSY}} = \frac{1}{(2\pi)^8\ell_p^9}\left(\int_{M} \Phi^{(\text{GI})} + \int_{\barM}\baranomaly\right)\, ,
\end{align}
where each integral picks out the relevant degree part of the polyform. 
Here, as usual, to define the Chern--Simons term in \eqref{D11action} one picks a twelve-manifold 
$\barM$ with $M=\partial\barM$ as boundary, extending also
the four-form $G$ \cite{Witten:1996md}. In the equivariant setting we must also extend
the Killing vector $K$ to $W$, and 
 $\Phi^G$ to an equivariantly closed form  on $\barM$ with associated
$\baranomaly$ given by the same formula \eqref{pair}. 

The key observation in \cite{BenettiGenolini:2026cdw} is that $(\Phi^{(\text{GI})},\baranomaly)$ on $(M,\barM)$ define a relative equivariant cohomology class, and the action 
\eqref{ISUSY} may be localized:
\begin{align}\label{main}
I_{\text{SUSY}}  =\frac{1}{(2\pi)^8\ell_p^9} \int_{\Sigmanew} \frac{\baranomaly}{e_{\barK}(N\Sigmanew)}\, .
\end{align}
Here $\Sigmanew\equiv\{\barK=0\}\subset \barM$ is the fixed point set of 
$\barK$ in $\barM$, with normal bundle $N\Sigmanew$ and associated equivariant Euler class $e_{\barK}$.   In particular, the boundary term one obtains when 
applying the Berline--Vergne--Atiyah--Bott (BVAB) formula \cite{BV:1982, Atiyah:1984px} precisely vanishes on using 
the first equation in~\eqref{pair}. 
Equation \eqref{main} shows that the supersymmetric supergravity action is computed via equivariant localization of the associated anomaly form.

So far our discussion has been classical, but quantum corrections 
play an important role in the above construction. 
The Chern--Simons form has the 
well-known 
one-loop correction
$C\wedge X_8$, 
where the eight-form $X_8=\frac{1}{192}(P_1^2-4P_2)$ is defined in terms of Pontryagin forms $P_1,P_2$ of the Riemann curvature two-form 
 \cite{Vafa:1995fj, Duff:1995wd}. Including this, together with a one-loop fermionic contribution 
from the gravitino, one can show 
that different choices 
of twelve-manifold $\barM$ and extension of $G$ 
to $\barM$ in \eqref{ISUSY} lead to $2\pi\ii \mskip2mu\Z$ shifts of $I$, so that the
quantum effective action $\ex^{-I}$ is well defined \cite{Witten:1996md}.  The second equation in \eqref{pair} should then be replaced by 
 \begin{align}\label{quantumanomaly}
\Phianom \mapsto  \frac{\ii}{6}(\Phi^G)^3 + \ii(2\pi \ell_p)^6 \Phi^G\wedge\Phi^{X_8}\, ,
\end{align}
where ${\Phi}{}^{X_8}$ is defined by 
replacing the Riemann 
curvature two-form $R^{ab}$ on $\barM$ by the covariant, equivariantly closed polyform
$R^{ab}-\frac{1}{2}(\diff {\barK}{}^\flat)^{ab}$, with $a,b=1,\ldots,12$ being tangent space indices.  The key formula \eqref{main} will still hold provided the corresponding corrected version of the first equation in \eqref{pair} holds, possibly up to the addition of a further $\diff_K$-exact term. We conjecture that this is true to all orders in $\ell_p$.
The full higher-derivative corrections to the M-theory effective action 
are not known, making this conjecture difficult to verify by a direct approach. However, 
$\Phianom$  in \eqref{quantumanomaly} is exact in $\ell_p$ because of the quantization argument. Therefore, if the conjecture is true, 
the expression for the supersymmetric action in \eqref{main} is an exact result, valid to all orders in $\ell_p$.

In deriving \eqref{main} we have assumed that $M$ is closed (compact without boundary).
For applications to holography  we will take $M$ to be a $Y_7$ fibration over $M_4$, 
with $Y_7$ a closed seven-manifold but $M_4$ a four-manifold with holographic 
conformal boundary $\partial M_4$ where the M2-brane field theory is defined. 
We assume that the required boundary terms to give a good variational principle,  supersymmetric 
holographic counterterms, and boundary BVAB term all cancel for supersymmetric configurations. 
There is  considerable 
evidence that this holds generally
 for equivariant supergravity localization \footnote{This has been proven rigorously in $D=4$, $\mathcal{N}=2$ gauged 
supergravity coupled to general vector multiplets in \cite{BenettiGenolini:2024lbj}. 
An alternative perspective is that equivariant localization defines 
a supersymmetric regularization scheme.}, with the results here
strengthening that evidence.   

\section{Grand potential}

We now apply \eqref{main} to the M-theory dual of ABJM theory at Chern--Simons level $k=1$, namely $M=EAdS_4\times S^7$ where $EAdS_4\cong\R^4$ denotes Euclidean anti-de Sitter spacetime with conformal boundary $S^3=\partial \R^4$. 

We take $\barM=EAdS_4\times \C^4\cong 
\C^6$, and the supersymmetric Killing vector to have the general diagonal 
form  $\barK = \sum_{i=1}^6 b_i\partial_{\varphi_i}$ where $\partial_{\varphi_i}$
 rotate $\C_i$ with weight~1. This identifies 
 $EAdS_4\cong \C^2_{b_1,b_2}$ where the ratio of weights
$b^2 \equiv {b_1}/{b_2}$
defines the squashing parameter of $S^3_b\cong \partial\C^2_{b_1,b_2}$ (the gravity dual was first constructed in \cite{Martelli:2011fu}). The vector $\barK$ has an isolated 
fixed point (a ``nut'') at the origin of $\C^{6}$. 
The Killing spinor $\epsilon$ is uncharged under $K$, which, with 
appropriate sign conventions, 
implies the constraint
\begin{align}\label{constraint}
\sum_{i=1}^6 b_i = 0 \quad \Leftrightarrow \quad  -\sum_{I=1}^4 u_I = b_1+b_2\, ,
\end{align}
where it is convenient to define $u_I\equiv b_{I+2}$, $I=1,\ldots,4$. 

In order to evaluate \eqref{main} using \eqref{quantumanomaly}  we need the zero-form part of $\Phi^{X_8}$ evaluated at an isolated fixed point:
\begin{align}\label{X80}
\Phi^{X_8}_{0} = \frac{1}{192 (2\pi)^4}\Big[\Big(\sum_{i=1}^6 
b_i^2\Big)^2 - 4\sum_{i<j} b_i^2b_j^2\Big]\, ,
\end{align}
where the $b_i$ arise as skew eigenvalues of $\diff {\barK}{}^\flat$. 
Next, crucially we identify $\mu$ with the flux of $G$ through $EAdS_4$:
\begin{align}\label{mudef}
\mu & \equiv \frac{\ii }{(2\pi)^2\ell_p^3}\int_{EAdS_4}G   \nonumber\\
& \equiv 
\frac{\ii }{(2\pi)^2\ell_p^3}\int_{\R^4}^{\text{eq}} {{\Phi}}{}^G
=  \frac{\ii }{b_1b_2\ell_p^3}{{\Phi}}{}^G_0\, .
\end{align}
Here notice that $\Phi^G$ is defined on the extension $\barM$, 
and generically has a zero-form component $\Phi^G_0$ in contrast to 
\eqref{phigee} on $M=\partial \barM$. 
In the first expression in \eqref{mudef} the integrand 
is
some dimensionless number 
times the volume form on $EAdS_4$. The second line regularizes this integral by 
turning it into an equivariant integral that picks out the zero-form part of the integrand at the origin, with measure the inverse equivariant Euler class $1/e_{K}(\R^4)=(2\pi)^2/b_1b_2$. 
Concretely, this means writing $\Phi^G=\diff_K \Phi^C$ on the boundary, 
where $\Phi^C$ can be interpreted as a holographic counterterm, 
and integrating the relative 
equivariant class $(-\Phi^C,\Phi^G)$ over $(S^3,\R^4)$, precisely as in \eqref{ISUSY}, 
which adds a boundary term to the first line of \eqref{mudef}. 

Using \eqref{X80} and \eqref{mudef}, 
the localized supersymmetric M-theory action \eqref{main} on $\barM=\C^6$ is 
\begin{align}\label{ISUSYfirst}
I_{\text{SUSY}} = -\frac{\ii (2\pi)^6}{\prod_{i=1}^6 b_i}\left[\frac{1}{6}\frac{(\ii b_1b_2\mu)^3}{(2\pi)^8} + \frac{(\ii b_1b_2\mu)}{ (2\pi)^2}\Phi^{X_8}_0\right]\, ,
\end{align}
where we have used the inverse equivariant Euler class $1/e_{\barK}(\C^6) = 
(2\pi)^6/\prod_{i=1}^6b_i$. We may further write
\begin{align}\label{ISUSYlocalized}
I_{\text{SUSY}} = &-\frac{1}{\newsigma_4}\Big[\frac{(b_1b_2)^2}{8\pi^2}\frac{\mu^3}{3}
 \nonumber \\ 
& \qquad \   + \frac{1}{24}\Big(\newsigma_4+ (b_1+b_2)\newsigma_3
 +{b_1b_2}\newsigma_2 \Big)\mu\Big]\, ,
\end{align}
where $\newsigma_r=\newsigma_r(\bm{u})$ are elementary symmetric polynomials
in the $u_I=b_{I+2}$, 
so that in particular $\newsigma_4=u_1 u_2 u_3 u_4$, and we have used the 
constraint \eqref{constraint}.

Changing to standard field theory variables via
\begin{align}\label{variablechange}
\Delta_I \equiv -\frac{2u_I}{b_1+b_2}\, ,\quad b \equiv \sqrt{b_1/b_2}\, , 
\end{align} 
so that $\sum_{I=1}^4\Delta_I = 2$, remarkably, we then find
\begin{align}\label{IisminusJ}
I_{\text{SUSY}} = -J(\mu,b,\bm{\Delta})\,,
\end{align}
where $J(\mu,b,\bm{\Delta})$ is the ABJM grand potential \eqref{grandpotential}, with
real masses/$R$-charge parameters $\bm{\Delta}=\{\Delta_I\}$ \footnote{See 
the formulas in section 2.1 of  \cite{Bobev:2025ltz}.}.  
In particular, notice that this computation identifies
\begin{align}\label{BtoX8}
B(b,\bm{\Delta}) = -e_K(\R^4)\int_{\C^6}^{\text{eq}}\Phi^{X_8} = -\frac{e_K(\R^4)}{e_K(\C^6)}\Phi^{X_8}_0\, .
\end{align}

\section{Change of ensemble}

The Airy conjecture in field theory is a statement about the grand potential of the localized matrix model \eqref{grandpotential}, which is a
cubic polynomial whose inverse Laplace transform gives
the Airy function. The dual gravitational computation
must be performed in the same ensemble: one should fix
the chemical potential $\mu$ in~\eqref{mudef} associated to the M-theory $C$-field, 
as we have done so far, rather than the M2-brane charge $N$. This point, which
has an analogue in four-dimensional gravity \cite{Hawking:1995ap}, has often
been obscured in holographic comparisons, as stressed  in
\cite{Gautason:2025plx}.  

We introduce 
\begin{align}
G_7&\equiv \pi_7 -\frac{1}{2}C\wedge G- (2\pi\ell_p)^6\omega_7\,,
\end{align}
where $\omega_7$ is a Chern--Simons form for $X_8=\diff\omega_7$ and $\pi_7$ is the canonical momentum associated with
the gauge-invariant part of the action, with $\pi_7=\ii*G+\dots$, where the dots refer to $\ell_p$ corrections. 
The M2-brane charge $N$ is then defined as the Page flux
\begin{align}
\label{eq:N_Page}
	N \equiv \frac{1}{(2\pi \ell_p)^6}\int_{Y_7}G_7 \, ,
\end{align}
with the equations 
of motion guaranteeing the conservation of $N$. 
 In order to hold fixed this flux  on the boundary $\partial M=\partial M_4\times Y_7$, rather than $C$, we add to the action the Legendre transform term \footnote{Finding a polarization of the M-theory phase space with fixed Page charges is generically non-trivial, as the Chern--Simons term implies that $G_7$ is not the momentum conjugate to $C$ and that the Page charges don't mutually commute. It requires choosing boundary data on the cohomology of the $C$-field, and a maximal commuting subset of Page charges \cite{Moore:2004jv}. At least for Freund--Rubin electric solutions, the first problem is moot, and the second problem becomes relevant only at the quantum level.}
\begin{align}
I_{\text{Legendre}} = \frac{\ii}{(2\pi)^8\ell_p^9}\int_{M}G\wedge G_7\, .
\end{align}
On the other hand, we may evaluate this integral by first integrating $G_7$ over $Y_7$, 
and then using \eqref{mudef}, which gives 
\begin{align}\label{Legendre}
I_{\text{Legendre}} = N \mu \, .
\end{align}

Thus, the M-theory  localized  supersymmetric quantum effective action, having changed 
ensemble to fix the M2-brane charge $N$, is given by 
\begin{align}
\label{eq:I_Canonical}
\exp(-I) & = \exp\left[-I_{\text{SUSY}} - I_{\text{Legendre}}\right] \nonumber \\
& = \exp \left[J(\mu,b,\bm{\Delta})-N \mu\right]\, ,
\end{align}
where we have used \eqref{IisminusJ}. We may interpret this as supersymmetric 
localization within the ``M-theory path integral'', where one should then integrate over the 
remaining parameter $\mu$ in the (equivariant) $G$-flux. 
This integral along the imaginary axis produces the inverse Laplace transform \eqref{Laplace}, 
up to the overall constant prefactor and exponentially small non-perturbative corrections. 
We interpret the former as due to the unknown measure, including one-loop corrections, 
in the path integral. 

\section{Toric Calabi--Yau four-folds}

Having derived the Airy function for ABJM theory at 
level $k=1$, 
we may straightforwardly generalize to $k>1$, and to more general Sasaki--Einstein internal 
spaces $Y_7$.

For $k>1$, using  results in \cite{Bobev:2025ltz}, we compute 
\begin{align}\label{Bkshift}
B(b,\bm{\Delta}) = -\frac{(2\pi)^4}{ku_1 u_2u_3u_4}\Phi^{X_8}_0 + \frac{k^2-1}{24k}\, ,
\end{align}
where the first term uses \eqref{BtoX8} for the orbifold geometry $\R^4\times (\C^4/\Z_k)$,
which simply contributes a $1/k$ factor. The additional term in \eqref{Bkshift} was 
already derived in \cite{Bergman:2009zh}, and accounts more carefully for the orbifold singularity 
contribution to the curvature, effectively modifying the quantization argument above \eqref{quantumanomaly}. We return to this point for $k=4$ below. 

More generally we can take $EAdS_4 \times Y_7$ with $Y_7$ any Sasaki--Einstein seven-manifold. 
For simplicity, we assume that $Y_7$ is toric, meaning $Y_7$ admits a $U(1)^4$ isometry, and we take $\barM=EAdS_4\times Z_8$, with $Z_8$ a smooth Calabi--Yau four-fold with boundary $\partial Z_8=Y_7$.
The Killing vector $\barK$ generates an isometry in $U(1)^4$ that we assume only has isolated fixed points in $Z_8$. 
We may label these $a=1,\ldots,\chi(Z_8)$,  with $\chi(Z_8)$ the Euler number. 
The Calabi--Yau condition ensures that the constraint 
\eqref{constraint} holds, where now the weights  $\bm{u}^a=\{u_I^a\}$ of $K$ will in general depend on the fixed point labelled by $a$.  

The localized supersymmetric action 
at each fixed point is again given by \eqref{ISUSYlocalized}, where in general $\mu_a$ 
can also now depend on the fixed point:
\begin{align}\label{ISUSYlocalizedtoric}
I_{\text{SUSY}}& = -\sum_{a=1}^{\chi(Z_8)}\frac{1}{\newsigma_4^a}\Big[\frac{(b_1b_2)^2}{8\pi^2}\frac{\mu_a^3}{3}
 \nonumber \\ 
&\qquad \ \   + \frac{1}{24}\Big(\newsigma_4^a+ (b_1+b_2)\newsigma_3^a  +{b_1b_2}\newsigma_2^a \Big)\mu_a\Big]\, ,
\end{align}
with $\newsigma_r^a=\newsigma_r(\bm{u}^a)$ elementary symmetric polynomials.

Recent work \cite{Cassia:2025jkr} proposed a geometric 
 formulation of the Airy function conjecture \eqref{partitionfunction} for M2-brane theories 
associated to toric Calabi--Yau four-folds, in terms of equivariant volumes and characteristic classes. We may 
derive their conjecture straightforwardly: introduce 
\begin{align}
C(\bm{u}) & \equiv \sum_{a=1}^{\chi(Z_8)} \frac{1}{\newsigma_4^a}\, ,\quad 
k_i(\bm{u}) \equiv \frac{1}{C(\bm{u})}\sum_{a=1}^{\chi(Z_8)} \frac{\newsigma_i^a}{\newsigma_4^a}\, , 
\end{align}
for $i=2,3$. 
Changing to field theory variables \eqref{variablechange}, and for simplicity setting 
$\mu_a=\mu$ for all $a=1,\ldots,\chi(Z_8)$ (a point we return to below), we find
\begin{align}
-I_{\text{SUSY}}& = \frac{2}{3\pi^2}\frac{C(\bm{\Delta})}{Q^4}\mu^3+\frac{\mu}{24}\chi(Z_8) \nonumber\\ 
& \quad +\frac{\mu}{12}C(\bm{\Delta})\left[\frac{2k_2(\bm{\Delta})}{Q^2}-k_3(\bm{\Delta})\right]\, ,
\end{align}
where $Q\equiv b+b^{-1}$. 
This precisely agrees with the  grand potential conjectured in \cite{Cassia:2025jkr}. 

We conclude this section with two comments. 
First, when $k=4$ we may resolve the $\C^4/\Z_k$ singularity to the toric Calabi--Yau four-fold 
$Z_8=\mathcal{O}(-4)\rightarrow \mathbb{CP}^3$. In this case one can check that \eqref{ISUSYlocalizedtoric} precisely reproduces \eqref{Bkshift}, \emph{including} 
the last  term. 
Second, we chose to set $\mu_a=\mu$ equal in 
\eqref{ISUSYlocalizedtoric}, but more generally 
these parameters include baryonic chemical potentials, and also 
quantized $G$-fluxes. For example, to see the latter 
 consider a four-cycle $D_4\subset Z_8$ that is mapped to itself under $\barK$.
The $G$-flux through $D_4$ can be localized and expressed
in terms of weights and the $\mu_a$ at fixed points in $D_4$. 
 We may interpret 
these as more general baryonic branches with $G$-flux. 

\section{Gravitational blocks}
\label{sec:Gravitational_Blocks}

It is straightforward to further extend our results from $EAdS_4$ to general four-dimensional 
spacetimes $M_4$, thus obtaining the holographic free energy of the SCFT, dual to $AdS_4\times Y_7$, when placed on $\partial M_4$.
Remarkably, there is no need for a  consistent Kaluza--Klein truncation on $Y_7$ to $D=4$ gauged supergravity, as we compute directly in eleven dimensions and localize. 
However, before continuing
we  pause to comment on the relation to 
recent 
work on $D=4$ conformal supergravity \cite{BenettiGenolini:2026qdm}.

Start with ABJM theory, and solve  the saddle point equation $\frac{\diff}{\diff \mu}(-I_{\text{SUSY}}-N \mu)=0$ for $\mu=\mu_*$, where $I_{\text{SUSY}}$ includes the last term in \eqref{Bkshift} for $k>1$. Defining 
\begin{align}\label{prepotential}
&\frac{16\pi^2}{b_1b_2} \mathcal{F}(\bm{\Delta},b_1,b_2) \equiv -\left.\left[I_{\text{SUSY}}+N\mu \right]\right|_{\mu=\mu_*}\nonumber\\
& \qquad = \frac{\pi Q^2\sqrt{2k \Delta_1\Delta_2\Delta_3\Delta_4}}{3}\left[N- B(b,\bm{\Delta})\right]^{3/2}\, ,
\end{align}
we find that $\mathcal{F}=\mathcal{F}(\bm{\Delta},b_1,b_2)$ agrees with the nut contribution arising from the
effective prepotential for $D=4$ conformal supergravity given in Eq. (16) of \cite{BenettiGenolini:2026qdm}, and conjectured in 
\cite{Hristov:2022lcw}. Note that reference  \cite{BenettiGenolini:2026qdm} also derived 
a fixed point formula for the on-shell action on $M_4$ of general topology 
and for a general homogeneous prepotential. Via \eqref{prepotential} these results capture the 
leading $z^{3/2}$ term in the expansion of $\log\text{Ai}(z)$. 
 
Continuing in the M2-brane ensemble with fixed $\mu$,  we now consider the fibration structure 
\begin{align}\label{fibre}
Y_7 \hookrightarrow M \stackrel{\pi}{\rightarrow} M_4
\end{align}
for the eleven-dimensional spacetime. We take $Y_7$ to be toric Sasaki--Einstein, as in the last section. For simplicity we assume that $\pi_*K\equiv\xi$ has isolated fixed points on $M_4$, 
labelled by an index $p=1,\ldots,\chi(M_4)$. Let the weights of $\xi$ on $TM_4|_p$ be $b_{p,1}, b_{p,2}$. We extend $M$ to $\barM$ by fibering $Z_8$ over $M_4$ 
via the associated bundle to \eqref{fibre}. The weights of 
$\barK$ on the fixed points in the fibres $Z_8$ over $p$ are $\bm{u}^a_p=\{u^a_{p,I}\}$. These satisfy the 
zero charge constraint
\begin{align}\label{constraint_Fibered}
\sigma_{p,1}b_{p,1}+\sigma_{p,2}b_{p,2} + \sum_{I=1}^4 u^a_{p,I} = 0\, ,
\end{align}
at each fixed point in $W$ labelled by $(p,a)$, where we 
have allowed for sign choices $\sigma_{p,i}\in\{\pm 1\}$, corresponding to 
spinor projections at each fixed point \footnote{On the other hand, the internal space is Calabi--Yau, with fixed-chirality spinors in each two-plane $\R^2\cong \C$.}. The same  localization calculations go through 
as in the last section. Assuming $\mu_p^a=\mu_p$ for simplicity (as above) and writing
$\bm{\mu}=\{\mu_p\}$ we obtain $I_{\text{SUSY}} = -J(\bm{\mu})$ with 
\begin{align}
J(\bm{\mu}) = \sum_{p=1}^{\chi(M_4)} J_p(\mu_p)\, , \quad  \mu_p \equiv \frac{\ii}{b_{p,1}b_{p,2}\ell_p^3}\Phi^G_0|_p\, .
\end{align}
 The action has  thus been written as a sum of ``gravitational blocks'' 
 \cite{Hosseini:2019iad} associated to each nut in $M_4$, with individual blocks 
\begin{align}
J_p(\mu_p) = \frac{C_p}{3}\mu_p^3+ B_p\mu_p\, ,
\end{align}
where
\begin{align}
B_p & = \frac{1}{24}\sum_{a=1}^{\chi(Z_8)} \frac{1}{e^a_{p,4}}\Big[e^a_{p,4} + 
(\hat{b}_{p,1}+\hat{b}_{p,2})e^a_{p,3} + \hat{b}_{p,1}\hat{b}_{p,2}e^a_{p,2}\Big]\, , \nonumber\\
C_p & = \frac{(b_{p,1}b_{p,2})^2}{8\pi^2}\sum_{a=1}^{\chi(Z_8)} \frac{1}{e^a_{p,4}} \, , 
\end{align}
$e^a_{p,r}=e_{r}(\bm{u}^a_p)$ are elementary symmetric polynomials, and we have defined $\hat{b}_{p,i}\equiv \sigma_{p,i}b_{p,i}$ (no sum on $i$). 
The same block structure arising from equivariant localization has been derived in lower-dimensional (gauged) supergravity  \cite{BenettiGenolini:2024hyd,BenettiGenolini:2026qdm}.

Finally, the Legendre transform term is  \eqref{Legendre}, where 
\begin{align}\label{summu}
\mu = \frac{\ii}{(2\pi)^2\ell_p^3}\int_{M_4}^{\text{eq}} \Phi^G = \sum_{p=1}^{\chi(M_4)} \mu_p\, .
\end{align}
While $\mu$ corresponds to boundary data, conjugate to $N$, the partially off-shell supersymmetric configurations
have an additional set of $\chi(M_4)-1$ parameters which do not. For the M-theory path integral,
these undetermined BPS parameters should be integrated over.
In particular, to obtain the M-theory path integral in the fixed $N$ ensemble
\footnote{In the fixed $\mu$ ensemble one should change variables from $\mu_p$ to $\mu$ and $\chi(M_4)-1$ relative variables. After integrating over the latter in a saddle point approximation, one finds $\log Z_\text{M-theory}(\mu)= \cC\mu^3/3+\cB\mu-\frac{1}{2}(\chi(M_4)-1)\log\mu+\dots$
for large $\mu$ . In particular, in our approach the $\log\mu$ term arises from integrating over the BPS locus and not directly from a one-loop determinant.} we thus  take
\begin{align}\label{keyresult}
Z_{\text{M-theory}} \propto \int \diff \mu_1\ldots \diff\mu_{\chi(M_4)} \exp\left[J(\bm{\mu}) -  N \mu \right]\, ,
\end{align}
where the integration contour 
over each $\mu_p$ is taken to be the imaginary axis. Notice that the  M2-brane charge $N$ defined in \eqref{eq:N_Page} does not depend on $p$: distinct fixed points in $M_4$ are joined by curves and $G_7$ is
closed on-shell, so Stokes' theorem equates the periods. Altogether,  this shows
 \begin{align}\label{MAiry}
Z_{\text{M-theory}} = \ex^{A}\prod_{p=1}^{\chi(M_4)}C_p^{-\frac{1}{3}}\text{Ai}
\left[\frac{(N-B_p)}{C_p^{\frac{1}{3}}}\right]\, ,
\end{align}
up to non-perturbative corrections. As commented earlier, the unknown measure/one-loop corrections 
are absorbed into the prefactor function $A$. Note  that the $\log N$ term in the expansion of this partition function is
\begin{align}
	\label{eq:logarithmic_corrections}
\log Z_{\text{M-theory}} \supset \sum_{p=1}^{\chi(M_4)} -\frac{1}{4}\log N = 
-\frac{\chi(M_4)}{4}\log N\, ,
\end{align}
thus confirming the conjectures in \cite{Bobev:2023dwx, Hristov:2021zai}, following work of \cite{Bhattacharyya:2012ye}. 
The 
approach of \cite{BenettiGenolini:2026qdm, Hristov:2022lcw} using \eqref{prepotential} does not capture this term.

As a simple application, we may use the general formula \eqref{MAiry} 
to recover the field theory result  $Z_{\text{BH}} \sim |Z_{S^3_b}|^2$ \cite{Bobev:2026lvl}, where the black hole partition function
is expressed as a product of two Airy functions.
We thus give a check of this result to all orders in \mbox{$1/N$}, which may be regarded as an $AdS$ generalization of the OSV conjecture \cite{Ooguri:2004zv}.

For brevity, we focus on ABJM theory, taking $Y_7 = S^7/\Z_k$.
For a black hole topology $M_4=\R^2\times S^2$, we have two fixed points  $p\in \{N,S\}$ at the poles of the horizon $S^2$. 
To compare with \cite{Bobev:2026lvl} we write
\begin{align}
	(b_{N,1}, b_{N,2}) & = (1,\omega) \,,\quad (b_{S,1}, b_{S,2}) = (1,-\omega) \,, \nonumber \\
	u_{p,I} &= -\frac{\Delta_{p,I}}{2}(\sigma_{p,1}b_{p,1}+\sigma_{p,2}b_{p,2}) \,,
\end{align}
with $\sum_{I=1}^4 \Delta_{p,I} = 2$.
Substituting into the above taking
\begin{align}
	& \text{SCI \ :}\quad \sigma_{N,1} = \sigma_{N,2} \,,\quad \sigma_{S,1} = - \sigma_{S,2} \,, \nn\\
	& \text{TTI \ :}\quad \sigma_{N,1} = \sigma_{N,2} \,,\quad \sigma_{S,1} = \sigma_{S,2} \, ,
\end{align}
for the superconformal index (SCI) and the topologically twisted index (TTI), respectively,
due to the distinct ways of preserving supersymmetry.
Then \eqref{MAiry} is in agreement 
%\note{}\footnote{In the simple case when $\Delta_{p,I}=1/2$, 
%for the SCI we can rewrite \eqref{keyresult} by changing variables $\mu=\mu_N+\mu_S$, $\nu=\mu_N-\mu_S$.
%After doing the integral over $\nu$ we obtain an expression as in Eq. (5.8) of \cite{Bobev:2026gir}, with an integral over $\mu$ with the same $\mu^{-1/2}$ measure, consistent with the $-\frac{1}{2}\log\mu$ term in the other ensemble.}
with \cite{Bobev:2026lvl} (see Eq.'s (4.17) and (5.18) of \cite{Bobev:2026lvl}).

%\JP{} The agreement with the TTI and SCI results above is at fixed $N$. We can make contact with the fixed $\mu$ results of \cite{Bobev:2026gir} for the SCI, in the special case where $\Delta_{p,I}=1/2$.
%In this case we have $C_N = C_S \equiv 4\mathcal{C}$ and $B_N = B_S \equiv \mathcal{B}$.
%Performing a change of variables
%$\mu=\mu_N+\mu_S$, $\nu=\mu_N-\mu_S$, we can integrate over the partially off-shell BPS moduli $\nu$ (while $\mu$ is fixed as boundary data), which in this case is a Gaussian integral. One then obtains $\log Z(\mu) = -\frac{1}{2}\log\mu + \frac{\mathcal{C}}{3} \mu^3 + \mathcal{B} \mu$, consistent with...
%
%\note{We also highlight that before going to the fixed $N$ ensemble, writing
%$\mu=\mu_N+\mu_S$, $\nu=\mu_N-\mu_S$ and evaluating the integral over $\nu$ in a saddle point approximation, in the fixed $\nu$ ensemble,
%for large $\mu$ one obtains t$\log Z(\mu) =+()\mu^3 +()\mu-\frac{1}{2}\log\mu$. For the special case when $\Delta_{p,I}=1/2$, this is consistent with
%the proposal of \cite{Bobev:2026gir}...}

\section{Discussion}

We have shown that the  grand potential \eqref{grandpotential}, the associated Airy partition function \eqref{partitionfunction}, and numerous related field theoretic observations, follow directly in gravity from relative equivariant localization of quantum M-theory. 
Previous attempts at deriving gravity corrections to large $N$ field theory results 
have used an effective four-dimensional approach -- for example,  see \cite{Dabholkar:2014wpa,  Dabholkar:2010uh, Bobev:2020egg,  Hristov:2021qsw, Hristov:2022lcw,  Bobev:2022eus, Hristov:2024cgj, BenettiGenolini:2026qdm}.  By contrast, 
we work in eleven dimensions and localize directly in an associated twelve-dimensional space.   
This automatically incorporates the full Kaluza--Klein tower that controls various corrections, which in a four-dimensional treatment must be supplied separately. We have no need 
for a consistent truncation.
Our derivation relies on \eqref{pair},\eqref{quantumanomaly}
holding at all orders, which is the main problem left open by our analysis; proving this in superspace building on \cite{Howe:2011tm,Soueres:2016qre,Galli:2026lsl} may be a fruitful approach.

There are many applications and extensions of our results to other 
 spacetimes, some of which we have alluded to already.  
These results open a direct route into perturbative quantum M-theory via equivariant localization.

\

\noindent \textit{Note added}: after this letter had been completed Ref. \cite{Bobev:2026gir} appeared, which discusses the ``$\mu$-ensemble'' for asymptotically 
$AdS_4\times S^7/\Z_k$ solutions.

\section*{Acknowledgments}
We thank Nikolay Bobev and Jesse van Muiden for discussions.
FG and JFS thank the Centro de ciencias de Benasque Pedro
Pascual for hospitality.
This work was supported in part by STFC grants ST/X000575/1 and
ST/X000761/1, and SNSF Ambizione grant PZ00P2\_208666.
JP is supported by a Dean's PhD studentship at Imperial College.
FG is supported by an STFC studentship.

\bibliography{biblio}{}

\end{document}